\documentclass[preprint]{aastex631}
\usepackage{amsmath}

\DeclareMathOperator{\sech}{sech}

\shorttitle{Particle Acceleration by Magnetic Reconnection in the Heliosheath}
\shortauthors{Seo et al.}

\begin{document}

\title{Can Magnetic Reconnection Accelerate Energetic Particles in the Heliosheath?}
\author[0000-0002-5550-8667]{Jeongbhin Seo}
\affiliation{Los Alamos National Laboratory, Los Alamos, NM 87545, USA}
\author[0000-0002-1250-4690]{Teresa Boland}
\affiliation{Los Alamos National Laboratory, Los Alamos, NM 87545, USA}
\affiliation{Department of Physics and Astronomy, Purdue University, West Lafayette, IN 47907, USA}
\author[0000-0003-4315-3755]{Fan Guo}
\affiliation{Los Alamos National Laboratory, Los Alamos, NM 87545, USA}
\affiliation{New Mexico Consortium, Los Alamos, NM 87544, USA}
\author[0000-0001-5278-8029]{Xiaocan Li}
\affiliation{Los Alamos National Laboratory, Los Alamos, NM 87545, USA}
\author[0000-0002-7836-7078]{Giulia Murtas}
\affiliation{Center for Kinetic Plasma Physics, West Virginia University, Morgantown, WV 26506, USA}

\begin{abstract}
The mechanism for accelerating energetic particles in the outer heliosphere is still under debate. The termination shock, turbulence, and magnetic reconnection have all been proposed as viable mechanisms. 
We investigate compression-driven particle acceleration during magnetic reconnection in the heliospheric current sheet using the MHD--SDE framework that incorporates energetic-particle feedback into the MHD evolution. 
For the heliosheath conditions considered here, we find that reconnection-driven acceleration is inefficient downstream of the termination shock, whereas reconnection near the heliopause can generate stronger compression and produce a power-law differential intensity spectrum $J(\varepsilon)\propto\varepsilon^{-5.4}$.
However, even the hardest spectrum achieved during magnetic reconnection remains much steeper than the typical $J(\varepsilon)\propto\varepsilon^{-1.5}$ spectrum observed in the heliosheath. 
This indicates that magnetic reconnection alone is insufficient to explain the acceleration of energetic particles in the heliosheath.
\end{abstract}


\section{Introduction}
\label{s1}
Energetic particle populations in the outer heliosphere, including pickup ions (PUIs) and anomalous cosmic rays (ACRs), have been observed from several keV to tens of MeV through both in-situ measurements and remote sensing \citep{Garcia-Munoz1973,Hovestadt1973,McDonald1974,Decker2005,Gloeckler2008,McComas2009,McComas2024}. The origin of these energetic particles is a long-standing problem.  Voyager measurements beyond the termination shock revealed persistent nonthermal energetic ion populations with power-law spectra extending from tens of keV to tens of MeV, with differential intensity approximately following $J(\varepsilon) \propto \varepsilon^{-1.5}$ \citep{Decker2005,Gloeckler2008}. The Interstellar Boundary Explorer (IBEX) identified a globally distributed energetic neutral atom (ENA) flux over energies of roughly 0.5–6 keV, providing remote evidence for heliosheath ions in this energy range \citep{Funsten2009,McComas2009,Desai2012,McComas2024}. The Interstellar Mapping and Acceleration Probe (IMAP) will extend ENA measurements with enhanced spectral and angular resolution, enabling unprecedented characterization of energetic particle distributions in the outer heliosphere \citep{McComas2018,Zirnstein2025imap,Reisenfeld2026}.

The heliospheric termination shock has long been proposed as the primary site of energetic particle acceleration in the outer heliosphere \citep{Pesses1981,Jokipii1986,Jokipii1992,Giacalone2010,Giacalone2025}. However, when Voyager 1 crossed the termination shock, the ACR intensity did not peak but continued to rise in the heliosheath, suggesting other acceleration mechanisms or a modified shock theory is needed~\citep{Stone2005,Stone2008,Cummings2008}. Later studies have considered that shock acceleration \citep{McComas2006,Guo2010,Kota2008,Senanayake2013,Cummings2024}, turbulence-driven acceleration \citep{Fisk2006,Fisk2012,Jokipii2010}, and magnetic reconnection \citep{Lazarian2009,drake2010,Drake2013,Zank2015} as possible mechanisms.

The sectored heliosheath region (see Figure~\ref{fig:schematic}) consists of alternating magnetic-polarity sectors separated by folded heliospheric current sheets (HCSs), formed as the ballerina-skirt-like HCS crosses the termination shock and is compressed and distorted by plasma compression and flow deflection in the heliosheath \citep{Burlaga2009,Czechowski2010,Richardson2015,AlonsoGuzman2026}.
As the sectored magnetic field piles up toward the heliopause, the spacing between neighboring sector boundaries is expected to decrease and become spatially nonuniform \citep{Czechowski2010,drake2010}.
The HCSs are possible sites for magnetic reconnection and energetic-particle acceleration.
\citet{Lazarian2009} proposed first-order Fermi acceleration in fast turbulent reconnection, where particles gain energy by bouncing between converging reconnecting magnetic fluxes in the sectored heliosheath.
\citet{drake2010,Drake2013} instead emphasized first-order Fermi acceleration in merging magnetic islands in the heliopause sector region.
However, it remains unclear whether magnetic reconnection under heliosheath conditions can produce the observed nonthermal ion spectra. In particular, the particle-in-cell (PIC) simulations of \citet{drake2010} failed to produce clear power-law energy spectra, making it difficult to compare theoretical predictions with observations. Over the past decade, a number of studies have shown that low-$\beta$ magnetic reconnection can effectively produce power-law energy spectra \citep{Li2018,li2019,Li2021,Zhang2021,Zhang2024,guo2015,guo2024,Oka2023,Drake2025}. It is also shown that flow compression is the main particle acceleration mechanism during magnetic reconnection, especially in the weak guide field regime \citep{Li2018pic,Du2018,Montag2017}. Applying the compression acceleration model to the near-Sun heliospheric current sheet HCS, recent studies have shown that magnetic reconnection typically produces a power-law spectrum $J(\varepsilon) \propto \varepsilon^{-3}$ or softer, consistent with observations by Parker Solar Probe \citep{Murtas2024,Murtas2026}. However, HCS acceleration in the heliosheath is in a very different plasma environment. In this paper, we revisit whether magnetic reconnection can accelerate particles to high energy in the heliosheath.
 
Global MHD models and Voyager-based studies indicate that plasma conditions vary significantly across the heliosheath, from high-$\beta$, PUI-dominated plasma downstream of the termination shock to more magnetically compressed, lower-$\beta$ plasma toward the heliopause \citep{drake2010,Dialynas2019,Adhikari2023,Bera2023}. 
\citet{Zenitani2020} has shown that plasma $\beta$ controls the plasma compressibility and reconnection outflow dynamics, and it is therefore  expected to strongly influence compression-driven particle acceleration. 
A goal of this study is to determine how reconnection-driven PUI acceleration changes across the plasma-$\beta$ conditions expected from the termination-shock region to the heliopause.

Previous fully kinetic PIC simulations provide insight into particle injection and reconnection microphysics and have been widely used to study particle acceleration in low-$\beta$ magnetic reconnection \citep{Oka2010, Li2017, Du2018, Guo2014, Guo2019, li2019, Li2021}. Recent hybrid-PIC simulations have begun to explore reconnection in regimes with higher $\beta$ \citep{Giai2024}. However, in high-$\beta$ plasmas like heliosheath, the magnetic free energy available per particle is relatively small, so capturing the development of a nonthermal tail can require longer integration times \citep{drake2010}. Applying PIC-based simulations directly to the heliosheath is difficult also because the characteristic sector spacing is on the order of several AU, and the typical PUI gyroradius is several orders of magnitude less.
Energetic particle transport equations provide a macroscopic description of energetic-particle transport and acceleration on MHD fluid scales, without requiring the gyroradius to be resolved \citep{parker1965,blanford1987,zank2014}.
This approach has been used to study compression-driven particle acceleration in fluid-scale magnetic reconnection regions \citep{Li2018,Li2022,seo2024,Seo2026,Murtas2024,Murtas2026}.
Here, we use the MHD--SDE framework developed in \citet{seo2024}, in which Parker's transport equation is solved using the stochastic differential equation (SDE) method \citep{Li2018,Li2022} and the resulting nonthermal particle pressure is coupled back to the MHD equations. 
This feedback is essential to consider as PUIs contribute a substantial pressure component in the heliosheath, similar to our earlier studies \citep{seo2024,Seo2026}.  
Several other macroscopic approaches also include energetic-particle feedback. The kglobal and MHD-gPIC model evolves energetic particles using guiding-center equations and couples their anisotropic stress back to the MHD plasma, allowing pressure anisotropy and the firehose constraint to regulate reconnection \citep{drake2019,arnold2021,Hu2026}. However, its original formulation does not include any spatial diffusion or pitch-angle scattering. The MHD-PIC method includes the Lorentz-force backreaction of energetic particles but requires their gyromotion scale to be resolved \citep{bai2015}.
We apply the MHD--SDE framework to model reconnection-driven particle acceleration across heliosheath conditions, from the downstream region of the termination shock to near the heliopause.
Our results show the hardest spectrum achieved during magnetic reconnection remains much steeper than the typical $J(\varepsilon)\propto\varepsilon^{-1.5}$ spectrum observed in the heliosheath, indicating that magnetic reconnection alone is insufficient to explain the acceleration of energetic particles in the heliosheath.

The paper is organized as follows. Section~\ref{s2} describes the numerical methods used in this study. The main results are presented in Section~\ref{s3}. A brief summary is given in Section~\ref{s5}.

\section{Method}
\label{s2}

\subsection{MHD--SDE Method}

In the MHD--SDE method \citep{seo2024}, we assume that the nonthermal particle distribution is nearly isotropic in momentum space, so that its feedback on the MHD flow can be described by an energetic-particle pressure. This approximation is supported by PIC simulations of weak-guide-field reconnection, which show nearly isotropic high-energy particle distributions \citep{Li2018pic,li2019}. In this study, this particle component corresponds to PUIs. 
Under these assumptions, the resistive MHD momentum and energy equations are given by
\begin{eqnarray}
    \frac{\partial\rho\boldsymbol{v}}{\partial t} 
    + \nabla\cdot\left(\rho \boldsymbol{v}\boldsymbol{v} + P^*\boldsymbol{I}-\boldsymbol{B}\boldsymbol{B}\right) 
    = -\nabla \cdot P_{\rm{PUI}}\boldsymbol{I}, 
    \label{meq}\\
    \frac{\partial E}{\partial t} 
    + \nabla\cdot\left[(E+P^*)\boldsymbol{v}-\boldsymbol{B}(\boldsymbol{B}\cdot \boldsymbol{v})\right] 
    = \nabla\cdot(\boldsymbol{B}\times\eta\boldsymbol{j})
    -\boldsymbol{v}\cdot(\nabla \cdot P_{\rm{PUI}}\boldsymbol{I}), 
    \label{Eeq}
\end{eqnarray}
where
\begin{eqnarray}
    E=P_g/(\gamma_g-1) 
    + (\rho \boldsymbol{v}\cdot\boldsymbol{v} + \boldsymbol{B}\cdot\boldsymbol{B})/2.
\end{eqnarray}
Here, $\rho$, $\boldsymbol{v}$, $\boldsymbol{B}$, $P_g$, $P_{\rm PUI}$, $\boldsymbol{j}$, $\gamma_g$, and $\eta$ represent the mass density, velocity, magnetic field, thermal pressure, PUI pressure, current density, adiabatic index of the thermal plasma, and resistivity, respectively. 
The total pressure is defined as $P^* = P_g + B^2/2$, which includes both the thermal and magnetic components. $\boldsymbol{I}$ is the unit tensor.

The pressure of PUI particles, $P_{\rm PUI}$, is obtained from the PUI distribution,
\begin{eqnarray}
P_{\rm PUI}
= \frac{1}{3}
\int u\,p\,
f(\boldsymbol{x}, p, t)\,
\mathrm{d}^{3}p,
\label{relP}
\end{eqnarray}
where $u$ is the speed of PUI particles. 
The distribution function, $f(\boldsymbol{x}, p, t)$, is a function of position, $\boldsymbol{x}$, specific momentum, $p$, and time, $t$. 

We solve Parker's transport equation for PUI particles to obtain their distribution:
\begin{equation}
\frac{\partial f}{\partial t}
+(\boldsymbol{v}+\boldsymbol{v_d})\cdot\nabla f
-\frac{1}{3}\nabla \cdot \boldsymbol{v} 
\frac{\partial f}{\partial \ln p} 
= \nabla\cdot(\boldsymbol{\kappa}\nabla f). 
\label{sde}
\end{equation}
Here, $\boldsymbol{v_d}$ represents the particle drift velocity, while $\boldsymbol{\kappa}$ denotes the spatial diffusion tensor. 
In our simulation, we only consider cases where $\boldsymbol{v_d}$ is negligible 
(e.g., $v_d \approx 0.05\ {\rm km\ s^{-1}}$ for 1 keV protons, a characteristic length scale of 1 AU, and a magnetic field of 0.1 nT \citep{isenberg1979}). 
We solve this equation using the SDE method, the details of which are described in \citet{Li2018}.

The spatial diffusion coefficient tensor is
\begin{equation}
\kappa_{ij} 
= \kappa_\perp \delta_{ij}
-\frac{(\kappa_\perp-\kappa_\parallel)\boldsymbol{B}_i\boldsymbol{B}_j}{\boldsymbol{B}^2},
\end{equation}
where $\delta_{ij}$ is the Kronecker delta, and $\kappa_\perp$ and $\kappa_\parallel$ are the perpendicular and parallel diffusion coefficients, respectively. 
According to quasi-linear theory, the parallel spatial diffusion coefficient for energetic protons in heliosheath turbulence can be estimated as \citep{giacalone1999, Murtas2024},
\[
\kappa_\parallel
=
\frac{u^3}{4L_c\Omega_0^2\sigma^2\gamma}
\csc\left(\frac{\pi}{\gamma}\right)
\left[
1+
\frac{8}{(2-\gamma)(4-\gamma)}
\left(\frac{\Omega_0 L_c}{u}\right)^\gamma
\right],
\]
where $L_c$ is the turbulence correlation length, $v$ is the particle speed, $\Omega_0=qB/(m_pc)$ is the proton gyrofrequency, and $\sigma^2=\langle\delta B^2\rangle/B_0^2$.
Here, $q$, $m_p$, and $c$ are the proton charge, proton mass, and speed of light, respectively, and $\gamma=5/3$ is used for a Kolmogorov-like turbulence spectrum.

In the present model, spatial diffusion is attributed only to the prescribed background turbulence. This assumption is consistent with previous three-dimensional reconnection simulations, which found self-generated magnetic fluctuations with $\langle\delta B^2\rangle/B_0^2$ of order $10^{-2}$--$10^{-1}$ in the fully developed turbulent state \citep{Huang2016}. Because this level is substantially below the variance of background heliosheath turbulence, $\sigma^2\sim1$ (see discussions below), we neglect diffusion associated with the reconnection-driven fluctuations.

\begin{deluxetable}{lcccccccc}
\tablewidth{\linewidth}
\tablecaption{Model parameters for the heliosheath cases considered in this study.
\label{tab:model_parameters}}
\tablehead{
\colhead{Case} 
& \colhead{$\beta$} 
& \colhead{$P_{\rm PUI,0}/P_0$} 
& \colhead{$B_0$ [nT]} 
& \colhead{$n_0$ [cm$^{-3}$]} 
& \colhead{$v_A$ [km s$^{-1}$]}
& \colhead{$L_0$ [AU]}
& \colhead{$\tau_A$ [yr]}
& \colhead{$\Delta_{\rm sec}$ [AU]}
}
\startdata
NTS  & 5   & 0.5 & 0.2 & $2.0\times10^{-3}$ & 98  & 25 & 1.22  & 5.0 \\
MID  & 1   & 0.3 & 0.4 & $2.0\times10^{-3}$ & 195 & 15 & 0.365 & 3.0 \\
HP   & 0.5 & 0.1 & 0.6 & $2.0\times10^{-3}$ & 293 & 10 & 0.162 & 2.0 \\
\hline
HP-W & 0.5 & 0.1 & 0.6 & $2.0\times10^{-3}$ & 293 & 10 & 0.162 & 3.3 \\
HP-C & 0.5 & 0.1 & 0.6 & $2.0\times10^{-3}$ & 293 & 10 & 0.162 & 1.1 \\
\enddata
\tablecomments{
NTS, MID, and HP denote the near-termination-shock, intermediate-heliosheath, and near-heliopause cases, respectively.
HP-W and HP-C denote the wide-spacing and compact-spacing HP cases, respectively.
The Alfv\'en crossing time is defined as
$\tau_A=L_0/v_A$.
The parameter $\Delta_{\rm sec}$ denotes the adopted sector spacing in the local simulation domain.
}
\end{deluxetable}

\subsection{Initial MHD Setup}

The models are initialized with uniform density and pressure and no bulk fluid motion. 
The resistivity is fixed at $\eta = 10^{-5}$.
The computational domain spans $x \in [-1,1]$ and $y \in [-0.5,0.5]$, resolved using a $2048 \times 1024$ grid. The Lundquist number $S = L_x v_A / (2\eta) = 10^{5}$, which places the system firmly in the high-Lundquist, plasmoid-unstable regime. Here, $v_A$ is the Alfv\'en speed. The realistic spatial scale for each case will be discussed in Section \ref{s.2.3}. 
All boundaries are taken to be periodic.

To capture a local patch of this sectored structure, we adopt a multiple current-sheet geometry, given by
\begin{equation}
\boldsymbol{B} = B_x(y)\, \hat{\boldsymbol{x}} + B_z(y)\, \hat{\boldsymbol{z}},
\end{equation}
where
\begin{equation}
B_x(y) = B_0 \left[ \sum_{i=1}^{N_s} (-1)^{i+1} 
\left( \tanh\left( \frac{y - y_i}{\lambda} \right) - 
\tanh\left( \frac{y + y_i}{\lambda} \right) \right) \right]-B_0,
\label{eq:Bx_sum_form}
\end{equation}
and
\begin{equation}
B_z(y) = B_0 \sum_{i=1}^{N_s} 
\left[ 
\sech\left( \frac{y - y_i}{\lambda} \right) 
+ \sech\left( \frac{y + y_i}{\lambda} \right) 
\right],
\end{equation}
with $y_i = (i-1/2)/N_s$. 
Here, $N_s$ controls the number of current-sheet pairs in the local domain and therefore sets the initial separation between neighboring current sheets. 
 $\lambda$ is half of the current-sheet thickness. 

We set $\lambda=0.005$, ensuring that each current sheet is resolved by ten grid cells. For the present simulations, the Lundquist number satisfies $S\gtrsim10^4$, placing the system in the plasmoid-unstable reconnection regime \citep{Bhattacharjee2009,Loureiro2016}. In this regime, variations in the initial current-sheet thickness mainly affect the onset time of reconnection, while the subsequent nonlinear evolution, including island formation and merging, is only weakly affected. Additional simulations with different initial sheet thicknesses (not shown here) also yield nearly unchanged saturated particle spectra, indicating that the main particle-acceleration results do not depend strongly on the choice of $\lambda$.
The thin current sheets, together with the particle noise introduced by nonthermal particles, provide perturbations that seed magnetic reconnection. 
Although the heliospheric current sheet is not expected to be globally force-free, we adopt a locally force-free configuration as an idealized initial equilibrium. 
This choice reduces artificial initial force imbalance and allows us to focus on how plasma $\beta$ and current-sheet separation regulate reconnection-driven compression and PUI acceleration.

\begin{figure}[ht!]
\includegraphics[width=0.5\textwidth]{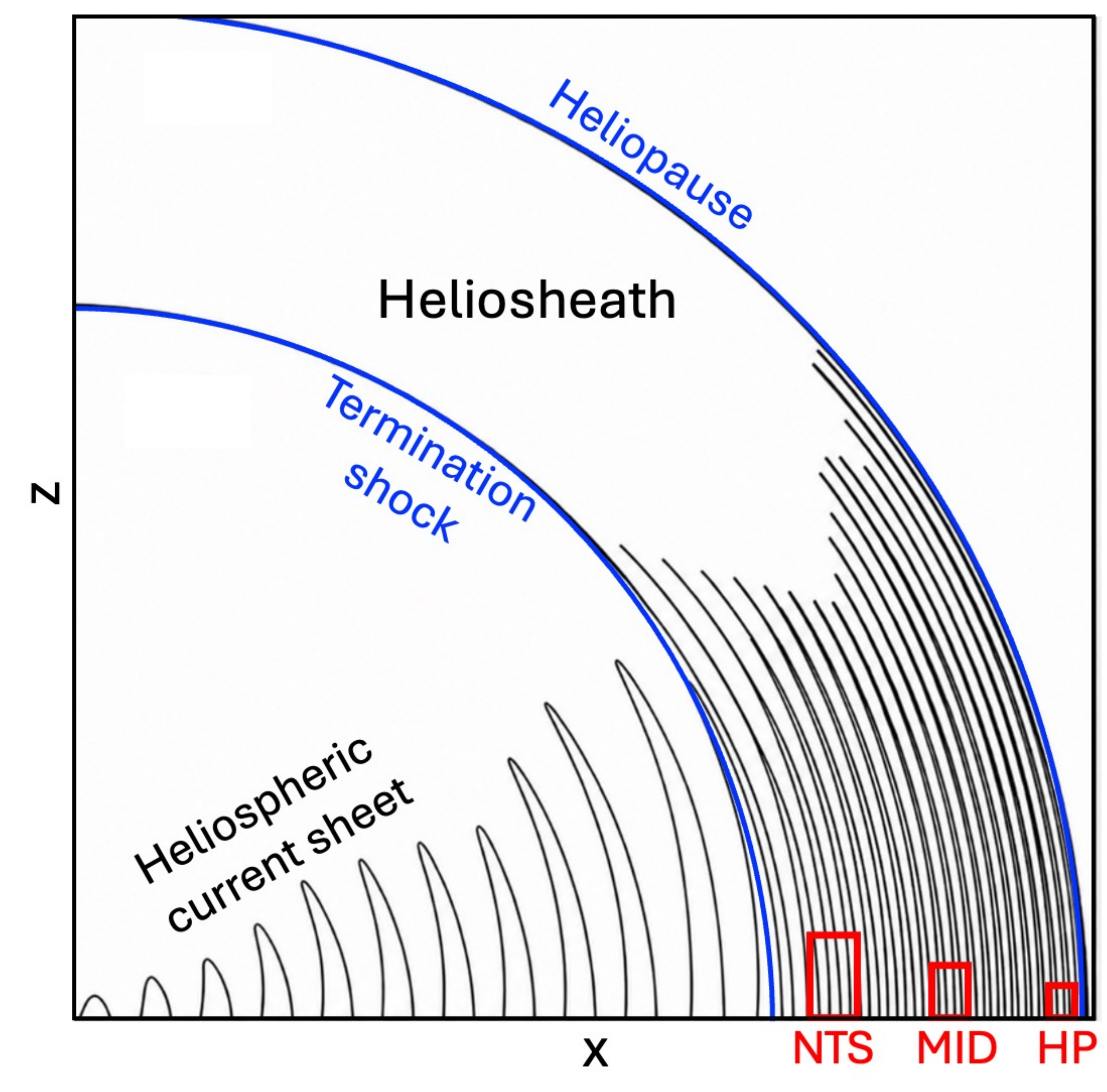}
\centering
\caption{Schematic illustration of the HCS structure, inspired by the heliosphere model of \citet{AlonsoGuzman2026}.
The red boxes indicate the local simulation regions used to represent the NTS, MID, and HP cases.
 \label{fig:schematic}}
\end{figure}

\subsection{Model Parameters and Normalization}
\label{s.2.3}
Motivated by previous global heliosphere simulations and Voyager observations 
\citep{Dialynas2019,Adhikari2023,Bera2023}, 
we construct a set of idealized local simulations that sample the transition from 
the high-$\beta$, PUI-rich plasma downstream of the termination shock to the 
more magnetically compressed, lower-$\beta$ plasma toward the heliopause.
As shown in Figure~\ref{fig:schematic}, we refer to these location-dependent cases as NTS, MID, and HP, representing 
near-termination-shock, intermediate-heliosheath, and near-heliopause conditions, respectively. 
We also include two additional HP-like cases, HP-W and HP-C, where W and C denote wide and compact current-sheet spacing, respectively, to examine how local sector spacing may affect reconnection-driven PUI acceleration.
The model parameters are summarized in Table~\ref{tab:model_parameters}.

The magnetic-field strength is taken to increase from $B_0=0.2~{\rm nT}$ in the NTS case to $B_0=0.6~{\rm nT}$ in the HP case, reflecting the expected field-strength enhancement toward the heliopause.
For simplicity, we use a common density, 
$n_0=2.0\times10^{-3}~{\rm cm^{-3}}$, 
motivated by Voyager~2 plasma observations just downstream of the termination shock \citep{Richardson2015}.

The initial thermal pressure is specified by the plasma beta through 
$P_0=\beta B_0^2/2$ in normalized units. Here, $\beta$ is defined using the thermal plasma pressure only.
Multi-component heliosheath models suggest that the PUI pressure fraction is largest immediately downstream of the termination shock and decreases toward the heliopause \citep{Bera2023}.
We therefore decrease $P_{\rm PUI,0}/P_0$ from the NTS case to the HP case, where, $P_{\rm PUI,0}$ denotes the initial PUI pressure.
Although the PUI pressure near the termination shock may be comparable to or larger than the thermal pressure in realistic heliosheath conditions, larger $P_{\rm PUI,0}$ can introduce substantial particle noise in our MHD--SDE simulations. So, for the NTS case, we set $P_{\rm PUI,0}=0.5P_0$, and use smaller values for the MID and HP cases to follow the expected decrease in PUI pressure toward the heliopause. A larger $P_{\rm PUI,0}$ would further reduce the compressibility \citep{seo2024}, so PUI acceleration in the NTS and MID case is expected to be even less efficient than in the present results.
Since $P_{\rm PUI,0}/P_0=0.1$--0.5 in our simulations, including the PUI pressure gives an effective plasma beta that is 10--50\% larger than the $\beta$ values.

We initialize a pre-existing PUI population to study its acceleration by magnetic reconnection.
The initial PUI energy is set to $\varepsilon_{\rm 0}=1~{\rm keV}$, which is comparable to the proton kinetic energy with a solar-wind speed of $\sim 400 {\rm km~s^{-1}}$.

We set the physical domain size to obtain the adopted characteristic sector spacing for each heliosheath region.
Following the sectored-heliosheath model of \citet{drake2010}, in which the sector spacing decreases from $4.7~{\rm AU}$ downstream of the termination shock to $1.9~{\rm AU}$ farther into the heliosheath, we prescribe fiducial sector spacings of $\Delta_{\rm sec}=5.0$, 3.0, and $2.0~{\rm AU}$ for the NTS, MID, and HP cases, respectively.
Because our fiducial setup contains four current sheets in a periodic domain with $L_0=5\Delta_{\rm sec}$, these choices correspond to simulation-box sizes of $L_0=25$, 15, and $10~{\rm AU}$.
So, the fiducial NTS, MID, and HP cases use $N_s=4$, while the HP-W and HP-C cases use $N_s=2$ and $N_s=8$, respectively, to produce wider and more compact current-sheet spacing.

For the particle diffusion coefficient, we estimate a characteristic heliosheath value using the quasi-linear expression given above.
Voyager~2 observations of magnetic turbulence in the heliosheath identify a characteristic scale $\ell\sim1.3~{\rm AU}$ separating the energy-injection and inertial-cascade regimes \citep{Fraternale2019}.
We take the turbulence correlation length as $L_c\sim1~{\rm AU}$.
For the NTS, MID, and HP parameter combinations considered here, we adopt a representative background turbulence level of $\sigma^2\sim1$, motivated by Voyager~1 observations in the heliosheath \citep{Zhao2025}. With this turbulence level, the quasi-linear estimate for $1~{\rm keV}$ protons gives $\kappa_{\parallel,0}\sim10^{-2}\kappa_0$, where $\kappa_0=L_0v_A$.
We also assume anisotropic diffusion with $\kappa_\perp=0.01\kappa_\parallel$ \citep{giacalone1999}.
We note that turbulence in the heliosheath may differ from that in other environments in terms of its compressibility and anisotropy, and its properties are not yet unambiguously constrained by observations \citep[e.g.,][]{Fraternale2019,Zhao2024}. Consequently, particle diffusion in the heliosheath may deviate from predictions based on standard quasi-linear theory \citep[e.g.,][]{Troskie2026}.

In the MHD-SDE model, the particle mean free path provides an additional constraint on the physical length normalization, so the domain cannot be rescaled independently while retaining the same particle transport conditions.

\begin{figure}[ht!]
\includegraphics[width=0.8\textwidth]{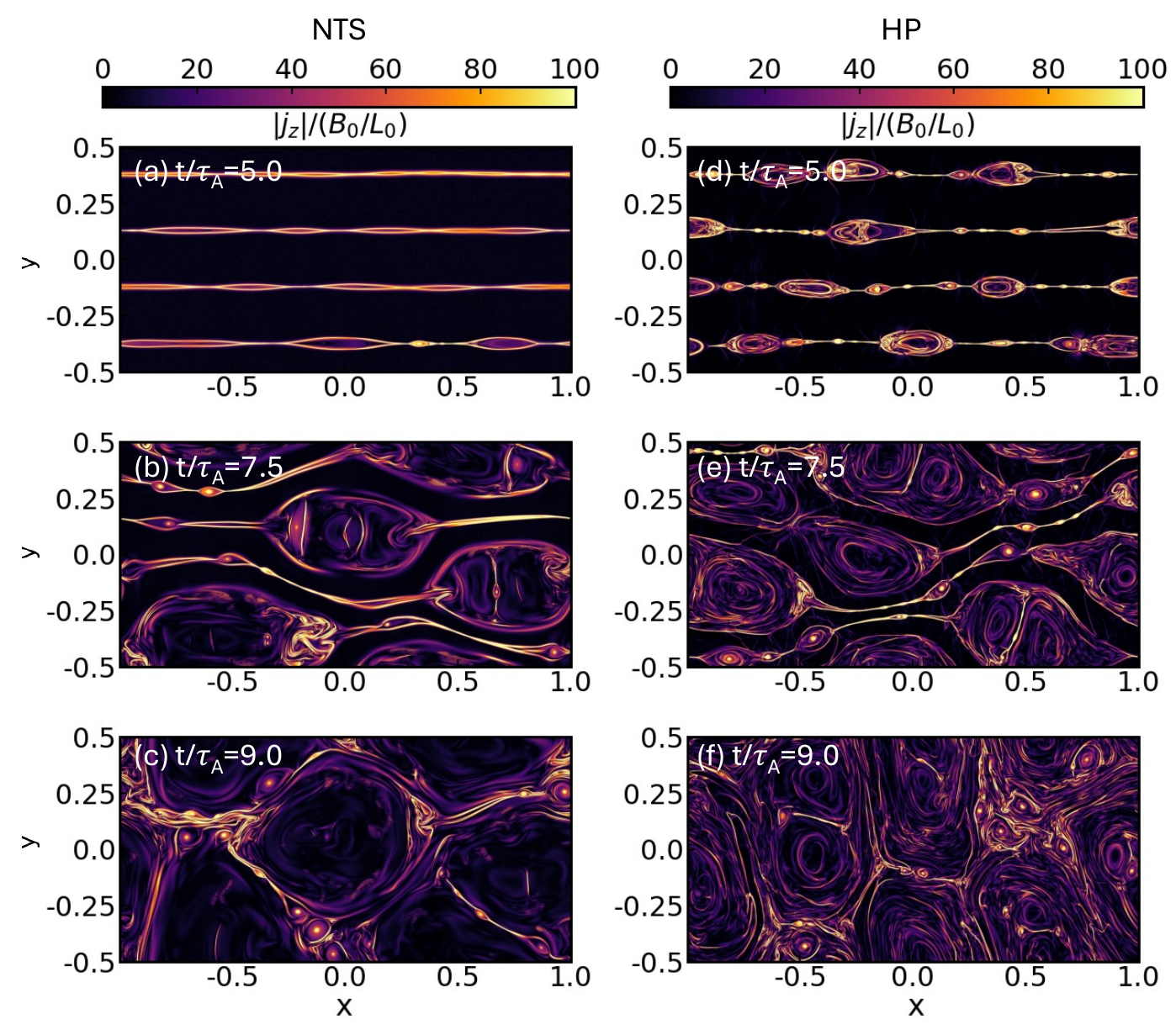}
\centering
\caption{Temporal evolution of the normalized out-of-plane current density $|j_z|$ for two heliosheath regions. Left column: NTS ($\beta=5$, $P_{\rm PUI,0}=0.5P_0$); right column: HP case ($\beta=0.5$, $P_{\rm PUI,0}=0.1P_0$). Panels (a)--(c) and (d)--(f) correspond to $t/\tau_A = 5.0$, $7.5$, and $9.0$, respectively.
 \label{fig:jz1}}
\end{figure}

\section{Results}
\label{s3}

A sufficiently thin current sheet can become unstable to the plasmoid instability, leading to the formation of multiple plasmoids \citep{loureiro2007,Bhattacharjee2009,Comisso2016}.
Figure~\ref{fig:jz1} presents the time evolution of the magnitude of the out-of-plane current density for the NTS and HP cases.
The nonlinear evolution depends on plasma $\beta$, which affects plasma compressibility during magnetic reconnection \citep{Zenitani2020}.
In the NTS case, the higher thermal and PUI pressures reduce plasma compressibility, so weaker current-sheet compression is associated with less pronounced small-scale plasmoid formation.
The resulting magnetic islands undergo less active merging and remain relatively smooth compared with those in the HP case.
In the lower-$\beta$ HP case, current-sheet fragmentation is more pronounced, producing a larger number of smaller plasmoids.
As discussed below, the HP case dissipates a much larger fraction of its magnetic energy than the NTS case, indicating more active reconnection-driven energy release.
At $t/\tau_A=5.0$ (Figure~\ref{fig:jz1}(a) and (d)), these morphological differences appear as fewer small-scale plasmoids in the NTS case and stronger plasmoid fragmentation in the HP case. 
Here, $\tau_A = L_0/v_A$ is the Alfvén crossing time.
At $t/\tau_A=7.5$ and $9.0$ (Figure~\ref{fig:jz1}(b), (c), (e), and (f)), both cases develop larger magnetic islands, but the HP case retains a more fragmented reconnection region, where enhanced small-scale plasmoid formation produces a larger number of smaller magnetic islands.

\begin{figure}[ht!]
\centering
\includegraphics[width=\textwidth]{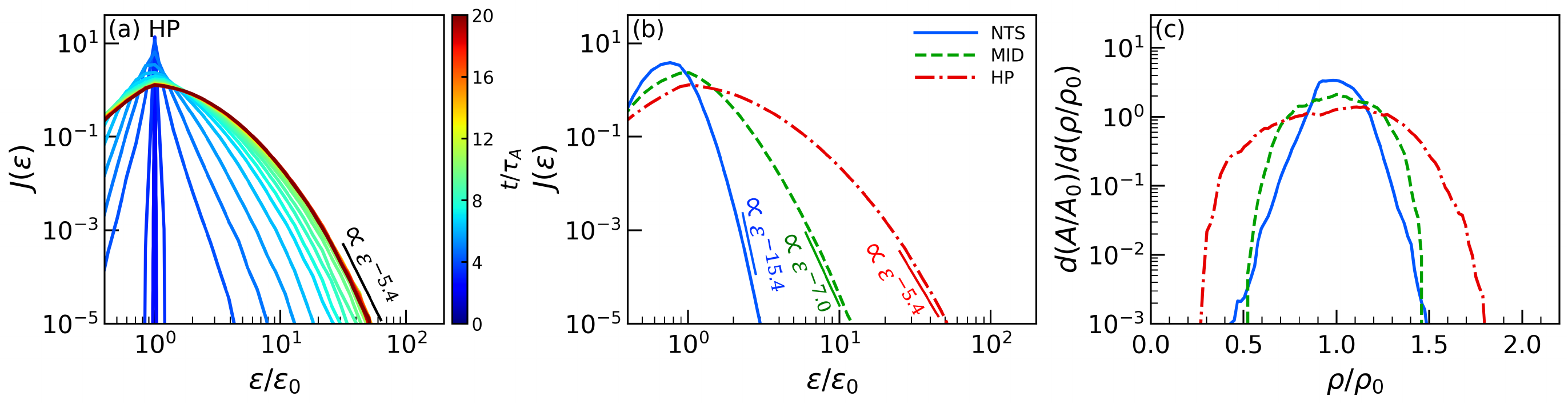}
\caption{
Particle spectra and density distribution for different heliosheath regions.
(a) Temporal evolution of the differential intensity $J(\varepsilon)$ in the fiducial HP case.
The color indicates time, $t/\tau_A$.
(b) Saturated PUI spectra for the NTS, MID, and HP cases. The blue solid, green dashed, and red dash-dotted lines denote the NTS, MID, and HP cases, respectively.
(c) Probability distribution functions of the normalized density, $\rho/\rho_0$, at $t/\tau_A=10$, using the same colors and line styles as in panel (b).
}
\label{fig:spec1}
\end{figure}

\begin{figure}[ht!]
\centering
\includegraphics[width=\textwidth]{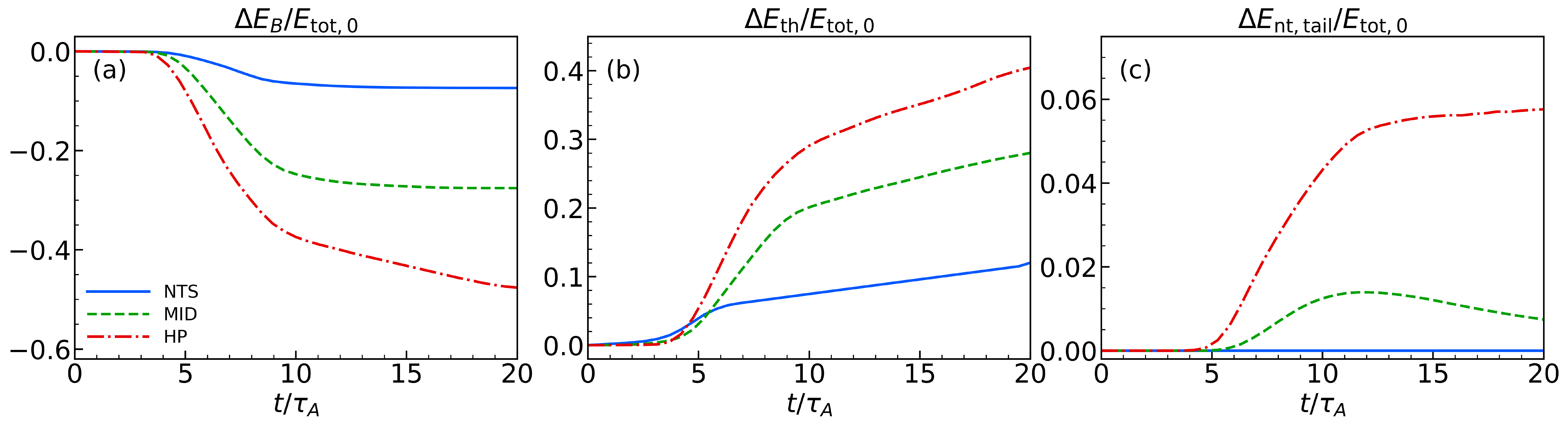}
\caption{
Temporal evolution of energy conversion for the NTS, MID, and HP cases. The blue solid, green dashed, and red dash-dotted lines correspond to the NTS, MID, and HP cases, respectively.
(a) Magnetic-energy change, $\Delta E_B/E_{\rm tot,0}$.
(b) Thermal-energy change, $\Delta E_{\rm th}/E_{\rm tot,0}$.
(c) Nonthermal tail-energy change, $\Delta E_{\rm nt,tail}/E_{\rm tot,0}$, where $E_{\rm nt,tail}$ is computed from particles with $\varepsilon/\varepsilon_0>2$.
}
\label{fig:E_var1}
\end{figure}

Figure~\ref{fig:spec1}(a) shows the temporal evolution of energetic particle spectrum $J(\varepsilon)$ in the fiducial HP case. 
As reconnection develops, a nonthermal tail forms and extends to higher energies, reaching a slope close to $J(\varepsilon)\propto\varepsilon^{-5.4}$. The particle spectra become approximately saturated by $t/\tau_A \approx 10$. Figure~\ref{fig:spec1}(b) compares the saturated spectra for the three plasma conditions, NTS, MID, and HP. The NTS case shows the steepest spectrum with only weak high-energy tail formation, while the MID case develops a nonthermal tail that remains steep.
The fiducial HP case produces a less steep spectrum than the NTS and MID cases. 
This relative hardening from NTS to MID to HP indicates that lower-$\beta$ plasmas provide more favorable conditions for compression-driven particle acceleration.
It is worth-noting that since the heliosheath plasmas move in a super-Alfvenic speed, the plasma residence time for each region may be shorter than the time needed for the acceleration becomes saturated. The plasma residence times can be estimated by dividing a 42 AU heliosheath thickness equally among the three regions and accounting for the progressively slower plasma flow toward the heliopause \citep{Dialynas2019,Krimigis2011}. We adopt representative flow speeds of 150, 75, and 25 km~s$^{-1}$ for the NTS, MID, and HP regions, respectively, which gives estimated residence times of approximately 0.4, 0.9, and 2.7 yr. The Alfv\'en speeds corresponding to the NTS, MID, and HP scalings are approximately 98, 195, and 293 km~s$^{-1}$, respectively, giving $10~\tau_A$ of approximately 12, 3.7, and 1.6 yr. A comparison of these timescales indicates that the saturated NTS and MID spectra require longer than the available local residence times. These spectra should therefore be regarded as upper limits, and accounting for finite advection would further reduce the acceleration efficiency in these regions. Since both cases show little particle acceleration even at saturation, this does not affect our main conclusion. In contrast, the HP case has sufficient time to approach saturation, but its energy spectrum is still too steep to explain the typical -1.5 slope observed in the heliosheath. 
Figure~\ref{fig:spec1}(c) shows the probability distribution functions of the normalized density, $\rho/\rho_0$, at $t/\tau_A=10$ which provide a measure of reconnection-driven compression. Here, $\rho_0$ denotes the initial thermal plasma density. This time step is chosen because it occurs shortly after the rapid magnetic-energy dissipation phase in all three cases, allowing the density distributions to capture the compression produced during the major reconnection stage (see Figure~\ref{fig:E_var1}(a)).
The NTS distribution remains narrowly concentrated around $\rho/\rho_0\simeq1$, indicating weak density compression.
The MID case shows a broader distribution, while the HP case extends to larger density contrasts. 
This ordering demonstrates that compressibility increases as $\beta$ decreases. 
Because magnetic tension drives stronger reconnection outflows in lower-$\beta$ plasmas, the resulting converging flows produce larger density enhancements and more favorable conditions for compression-driven particle acceleration.
To quantify this trend, we compute the RMS density fluctuation, $\delta\rho_{\rm RMS}/\rho_0$, from the PDFs presented in Figure~\ref{fig:spec1}(c).
For the HP, MID, and NTS cases, we obtain 
$\delta\rho_{\rm RMS}/\rho_0 \simeq 0.275$, 0.176, and 0.113, respectively. 
These values decrease with increasing $\beta$, consistent with the expected reduction of plasma compressibility at higher $\beta$. 
This trend is consistent with the $\beta$-dependent reconnection behavior discussed in \citet{Zenitani2020}.

We next examine the energy evolution of the NTS, MID, and HP cases to determine how much magnetic energy is released by reconnection in each heliosheath sector region and how this energy conversion is associated with plasma heating and PUI acceleration.
Figure~\ref{fig:E_var1} shows the temporal evolution of magnetic, thermal, and nonthermal PUI tail energies normalized by the initial total energy, $E_{\rm tot,0}$, for NTS, MID, and HP cases. 
Here, the nonthermal PUI tail energy, $E_{\rm nt,tail}$, is computed from PUI particles with $\varepsilon/\varepsilon_0>2$. 
As shown in Figure~\ref{fig:E_var1}(a), the magnetic-energy release is weakest in the NTS case and becomes progressively stronger from the MID to HP cases. 
The rapid decrease in magnetic energy marks the major reconnection stage, during which magnetic energy is efficiently released.
After this rapid-dissipation phase, the evolution becomes more turbulent, similar to the states shown in Figures~\ref{fig:jz1}(c) and (f), and the remaining magnetic energy is dissipated more gradually.
The corresponding thermal-energy evolution is shown in Figure~\ref{fig:E_var1}(b). 
The thermal-energy increase follows the same overall ordering as the magnetic-energy decrease, indicating that a large fraction of the released magnetic energy is converted into plasma heating.

\begin{figure}[ht!]
\centering
\includegraphics[width=0.9\textwidth]{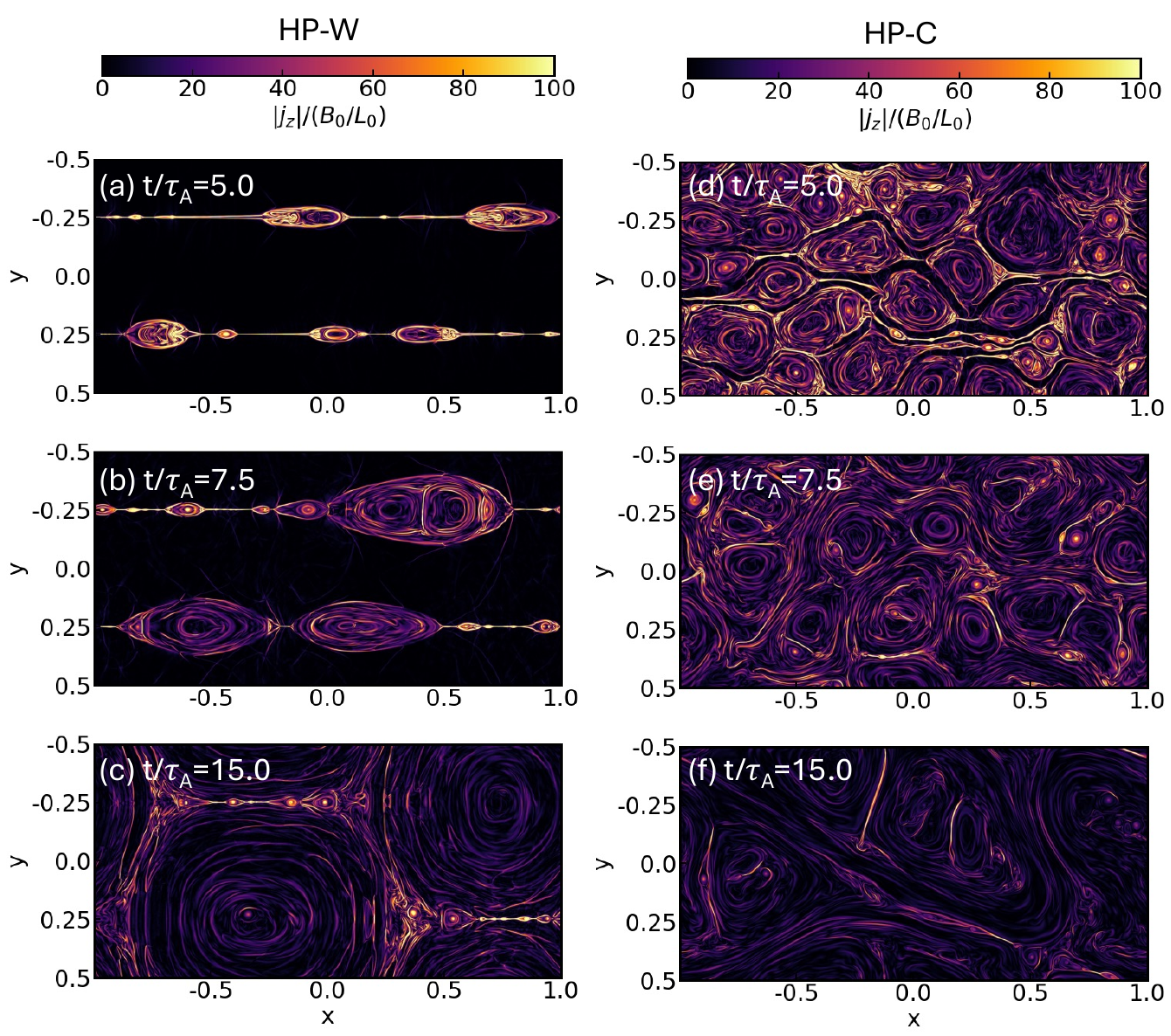}
\caption{
Time evolution of the normalized out-of-plane current density, 
$|j_z|/(B_0/L_0)$, for the HP cases with different current-sheet spacings. 
Panels (a)--(c) show the HP-W case at $t/\tau_A=5.0$, 7.5, and 15.0, respectively, 
while panels (d)--(f) show the HP-C case at the same times.
}
\label{fig:jz2}
\end{figure}

Figure~\ref{fig:E_var1}(c) shows the evolution of the nonthermal PUI tail energy. 
The NTS case produces almost no nonthermal PUI tail, while the MID case shows only weak tail growth. 
The HP case, by contrast, produces a much larger nonthermal component, demonstrating that the lower-$\beta$ HP-like regime provides more favorable conditions for compression-driven particle acceleration. 
In the MID case, the nonthermal tail energy decreases after reaching a modest peak, suggesting that particle diffusion into expanding regions leads to adiabatic cooling associated with $\nabla\cdot\boldsymbol{v}>0$.

Overall, the energy evolution shows that lower-$\beta$ conditions allow reconnection to release a larger fraction of magnetic energy, produce more efficient plasma heating, and drive stronger PUI acceleration.

We next discuss how the current-sheet spacing affects particle acceleration. As a reminder, HP-W and HP-C refer to the HP cases with wider and more compact current-sheet spacings, respectively.
Figures~\ref{fig:jz2}(a)--(c) show that the HP-W case evolves relatively independently along each initial current sheet and still retains coherent high-$|j_z|$ layers at $t/\tau_A=15.0$, even after magnetic islands from neighboring sheets become close to one another.
In contrast, Figures~\ref{fig:jz2}(d)--(f) show that the HP-C case undergoes much earlier interactions between neighboring current sheets, with many islands already formed at $t/\tau_A=5.0$, active cross-sheet island merging by $t/\tau_A=7.5$, and a highly turbulent state by $t/\tau_A=15.0$, where little trace of the initial current-sheet structure and few strong high-$|j_z|$ current sheets remain.

\begin{figure}[ht!]
\centering
\includegraphics[width=0.8\textwidth]{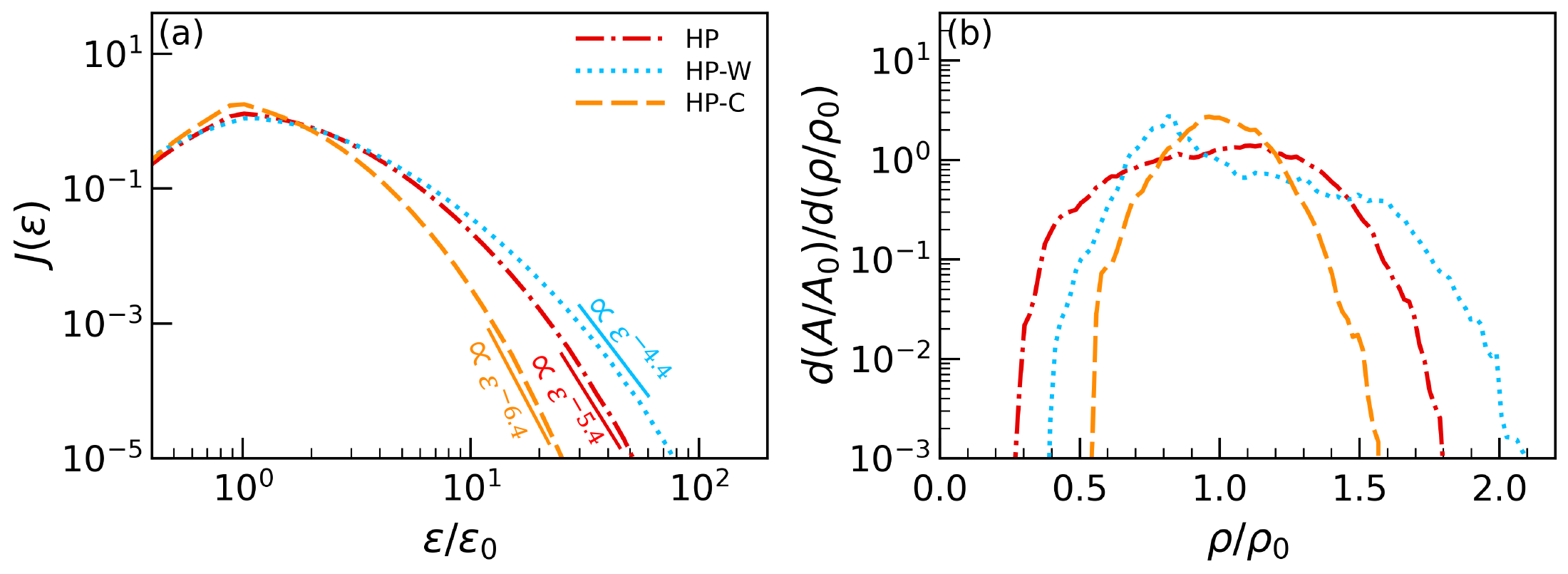}
\caption{
Particle spectra and density distribution for the HP cases with different current-sheet spacings.
(a) Saturated PUI spectra for the HP, HP-W, and HP-C cases. The red dash-dotted, cyan dotted, and orange dashed lines denote the HP, HP-W, and HP-C cases, respectively.
(b) Probability distribution functions of the normalized density, $\rho/\rho_0$, at $t/\tau_A=10$, using the same colors and line styles as in panel (a).
}
\label{fig:spec2}
\end{figure}

As shown in Figure~\ref{fig:spec2}(a), the HP-W case produces a somewhat harder high-energy spectrum than the fiducial HP case, with a slope close to $J(\varepsilon)\propto\varepsilon^{-4.4}$, whereas the HP-C case produces a steeper spectrum, with a slope close to $J(\varepsilon)\propto\varepsilon^{-6.4}$.

This ordering is consistent with the compression trend inferred from the density distributions in Figure~\ref{fig:spec2}(b). The stronger extension of the HP-W density distribution toward larger values in the $\rho/\rho_0 > 1$ range indicates stronger localized compression, while the $\rho/\rho_0 < 1$ region arises from mass conservation. The shape of the HP-W density PDF is similar to that found in the single-current-sheet reconnection simulations of \citet{Li2018}. In contrast, the HP and HP-C cases have already undergone stronger interactions between neighboring current sheets and developed into a more turbulent state by $t/\tau_A = 10$, resulting in density PDFs that are more symmetric around $\rho/\rho_0 = 1$. The density distribution in the HP-C case is consistent with weaker localized compression caused by earlier interaction and fragmentation of neighboring current sheets.

Although not shown here, the HP-C case dissipates magnetic energy earlier than the HP-W case, with the released energy contributing more to plasma heating than to PUI acceleration.
Thus, current-sheet spacing can modulate localized compression and PUI energization, but plasma $\beta$ remains the more fundamental control parameter because it affects both magnetic-energy release and plasma compressibility.

To conclude, near the termination shock, magnetic reconnection in the HCS produces only weak compression and inefficient particle acceleration. 
Toward the heliopause, reconnection can generate stronger compression under lower-$\beta$ conditions, leading to more efficient particle energization. 
Even in that case, however, the resulting differential intensity spectra across different sector spacings have power-law slopes of roughly $-4$ to $-6$, substantially steeper than the observed slope of $-1.5$.
If the simulated and observed spectra are normalized at $\varepsilon=1~{\rm keV}$, the discrepancy at $100~{\rm keV}$ remains at least about six orders of magnitude, even for the hardest simulated spectrum.
This indicates that compression-driven reconnection alone is insufficient to explain the energetic particle population.  
In the more magnetically dominated solar flare conditions, \citet{Li2018} obtained a proton differential intensity $J$ slope close to $-1.5$ for $\beta = 0.1$, with a diffusion coefficient approximately one order of magnitude smaller than that used here and without energetic-particle feedback. These conditions are substantially more favorable for acceleration than those expected in the heliosheath. Therefore, other mechanisms are required to explain the observed energetic particles spectra in the heliosheath.

\section{SUMMARY}
\label{s5}
We have investigated compression-driven particle acceleration during magnetic reconnection in the heliospheric current sheet using the MHD--SDE framework that couples large-scale MHD evolution with energetic-particle transport and feedback \citep{seo2024}. 
We considered three heliosheath conditions from the near-termination-shock region to the heliopause to quantify how reconnection-driven particle acceleration changes across the heliosheath.
Near the termination shock, where the plasma $\beta$ is high and pickup-ion pressure provides a substantial contribution to the total pressure, reconnection produces weak compression and inefficient particle acceleration. 
The resulting particle spectrum does not develop a clear nonthermal tail. 
Toward the heliopause, where magnetic pressure becomes more important relative to the thermal and pickup-ion pressures, reconnection produces stronger magnetic-energy release, enhanced plasma heating, and more pronounced compressive structures. 
The lower-$\beta$ HP-like condition therefore provides the most favorable environment for compression-driven particle energization among the location-dependent cases considered here. 
Nevertheless, even in this case, the resulting particle spectrum remains much steeper than the observed heliosheath energetic-particle spectrum, indicating that reconnection alone cannot reproduce the observed spectrum.
Current-sheet spacing produces secondary variations in particle energization near the heliopause, but the dominant trend is controlled by the plasma-$\beta$ variation across the heliosheath.

Taken together, our results suggest that reconnection can produce localized particle energization under lower-$\beta$ HP-like conditions, but the resulting spectra remain too steep to account for the observed heliosheath energetic-particle spectra, likely requiring additional acceleration mechanisms.

\section*{Acknowledgements}
We acknowledge the support from Los Alamos National Laboratory through the LDRD program, DOE OFES, and NASA programs through grant NNH17AE68I,
80HQTR20T0073, and 80HQTR21T0005, 80HQTR21T0117, NNH230B17A, and NNH240B72A. J.S. acknowledges the support from the Los Alamos National Laboratory (LANL) through its Center for Space and Earth Science (CSES). CSES is funded by LANL’s Laboratory Directed Research and Development (LDRD) program under project number 20240477CR. F. G. acknowledges NSF Award No. 2334932. X. L. acknowledges the support from NASA grant 80NSSC21K1313, National Science Foundation grant No. AST-2107745 and AGS-2334930, University of California, Berkeley through subcontract No. 00011753, and Smithsonian Astrophysical Observatory through subcontract No. SV1-21012. 
The simulations used resources provided by the Los Alamos National Laboratory Institutional Computing
Program, which is supported by the U.S. Department of Energy National Nuclear Security Administration under Contract No. 89233218CNA000001, and the National Energy Research Scientific Computing Center (NERSC), a Department of Energy Office of Science User Facility using NERSC award FES-ERCAP0031616. 

\bibliography{refs}{}
\bibliographystyle{aasjournal}

\end{document}